\documentclass[preprint2,11pt]{aastex}

\usepackage{natbib}
\bibpunct{(}{)}{,}{n}{}{,}
\citeindextrue

\def\HST{{\it HST}}
\def\FUSE{{\it FUSE}}
\def\HUT{{\it HUT}}

\def\arcsec{\ifmmode '' \else $''$\fi}
\def\arcmin{\ifmmode ' \else $'$\fi}
\def\arcsecpoint{\ifmmode ''\!. \else $''\!.$\fi}
\def\arcminpoint{\ifmmode '\!. \else $'\!.$\fi}
\def\cc{\ifmmode {\rm cm}^{-3} \else cm$^{-3}$\fi}
\def\cl{\ifmmode {\rm cm}^{-2} \else cm$^{-2}$\fi}
\def\micron{\ifmmode \mu{\rm m} \else $\mu$m\fi}
\def\kms{\ifmmode {\rm km\,s}^{-1} \else km\,s$^{-1}$\fi}
\def\Hubble{\ifmmode {\rm km\,s}^{-1}\,{\rm Mpc}^{-1}
        \else km\,s$^{-1}$\,Mpc$^{-1}$\fi}
\def\ergsec{\ifmmode {\rm ergs\;s}^{-1} \else ergs s$^{-1}$\fi}
\def\ergscm{\ifmmode {\rm ergs\,s}^{-1}\,{\rm cm}^{-2}
          \else ergs\,s$^{-1}$\,cm$^{-2}$\fi}
\def\ergscmA{\ifmmode {\rm ergs\,s}^{-1}\,{\rm cm}^{-2}\,{\rm \AA}^{-1}
          \else ergs\,s$^{-1}$\,cm$^{-2}$\,\AA$^{-1}$\fi}
\def\ergscmHz{\ifmmode {\rm ergs\,s}^{-1}\,{\rm cm}^{-2}\,{\rm Hz}^{-1}
          \else ergs\,s$^{-1}$\,cm$^{-2}$\,Hz$^{-1}$\fi}
\def\Msun{\ifmmode M_{\odot} \else $M_{\odot}$\fi}
\def\Lsun{\ifmmode L_{\odot} \else $L_{\odot}$\fi}
\def\qo{\ifmmode q_{0} \else $q_{0}$\fi}
\def\Ho{\ifmmode H_{0} \else $H_{0}$\fi}
\def\gtsim{\raisebox{-.5ex}{$\;\stackrel{>}{\sim}\;$}}
\def\lya{Ly$\alpha$}

\def\cii{C\,{\sc ii}}

\newcommand{\hi}{{\sc H~i}}
\newcommand{\heii}{He~{\sc ii}}

\def\eps@scaling{.95}
\def\epsscale#1{\gdef\eps@scaling{#1}}
\def\plotone#1{\centering \leavevmode
\epsfxsize=\eps@scaling\columnwidth \epsfbox{#1}}

\def\plotfiddle#1#2#3#4#5#6#7{\centering \leavevmode
\vbox to#2{\rule{0pt}{#2}}
\includegraphics{#1}}
\slugcomment{Version 2.5, June 29, 2001.}

\shorttitle{FUSE Observations of Ionized Helium in the IGM}
\shortauthors{Kriss et al.}

\begin{document}

\onecolumn
\title{Resolving the Structure of Ionized Helium in the Intergalactic Medium
with the Far Ultraviolet Spectroscopic Explorer}

\author{
G. A. Kriss\altaffilmark{1,2,*},
J. M. Shull\altaffilmark{3},
W. Oegerle\altaffilmark{4},
W. Zheng\altaffilmark{2},
A. F. Davidsen\altaffilmark{2},
A. Songaila\altaffilmark{5},
J. Tumlinson\altaffilmark{3},
L. L. Cowie\altaffilmark{5},
J.-M. Deharveng\altaffilmark{6},
S. D. Friedman\altaffilmark{2},
M. L. Giroux\altaffilmark{3},
R. F. Green\altaffilmark{7},
J. B. Hutchings\altaffilmark{8},
E. B. Jenkins\altaffilmark{9},
J. W. Kruk\altaffilmark{2},
H. W. Moos\altaffilmark{2},
D. C. Morton\altaffilmark{8},
K. R. Sembach\altaffilmark{2},
T. M. Tripp\altaffilmark{9}
}

\altaffiltext{1}{Space Telescope Science Institute,
        3700 San Martin Drive, Baltimore, MD 21218}
\altaffiltext{2}{Center for Astrophysical Sciences, Department of Physics and
        Astronomy, The Johns Hopkins University, Baltimore, MD 21218--2686}
\altaffiltext{3}{CASA and JILA, Department of
        Astrophysical and Planetary Sciences, University of Colorado,
        Campus Box 389, Boulder, CO 80309}
\altaffiltext{4}{Laboratory for Astronomy and Solar Physics, Code 681,
        NASA/Goddard Space Flight Center, Greenbelt, MD 20771}
\altaffiltext{5}{University of Hawaii, Institute for Astronomy, 
	2680 Woodlawn Road, Honolulu, HI 96822}
\altaffiltext{6}{Laboratorie d'Astronomie Spatiale, BP 8,
	13376 Marseille Cedex 12, France}
\altaffiltext{7}{Kitt Peak National Observatory, National Optical Astronomy Observatories, P. O. Box 26732,
        950 North Cherry Ave., Tucson, AZ, 85726-6732}
\altaffiltext{8}{Herzberg Institute of Astrophysics, National Research Council
        of Canada, Victoria, BC, V8X 4M6, Canada}
\altaffiltext{9}{Princeton University Observatory,
	Princeton, NJ 08544\\
\ \ \ \ \ \ \ \ \ $^*$
To whom correspondence should be addressed. E-mail: gak@stsci.edu }

\begin{abstract}
The neutral hydrogen (\hi) and the ionized helium (\heii) absorption
in the spectra of quasars are unique probes of structure in the
early universe.  We present {\it Far-Ultraviolet Spectroscopic Explorer}
observations of the line of sight to the quasar HE2347--4342 in the
1000--1187 {\AA}ngstrom band at a resolving power of 15,000.
We resolve the \heii\ \lya\ absorption as a discrete forest of absorption lines
in the redshift range 2.3 to 2.7.
About 50 percent of these features have \hi\ counterparts
with column densities $N_{HI} > 10^{12.3}$ per square centimeter
that account for most of the observed opacity in \heii\ \lya.
The \heii\ to \hi\ column density ratio ranges from 1 to $>$1000 with an
average of $\sim$80. Ratios of $<$100 are consistent with photoionization of the
absorbing gas by a hard ionizing spectrum resulting from the integrated light
of quasars, but ratios of $>$100 in many locations indicate additional
contributions from starburst galaxies or heavily filtered quasar radiation.
The presence of \heii\ \lya\ absorbers with no \hi\ counterparts indicates
that structure is present even in low-density regions, consistent with
theoretical predictions of structure formation through gravitational
instability.
\end{abstract}


\parbox{5.7in}{
\vskip 0.4in
}


\twocolumn

The intergalactic medium (IGM) is the gaseous reservoir that provides the
raw material for the galaxies that dominate our view of the visible universe.
By observing distant bright objects such as quasars, we can explore the IGM
by examining the absorption features it imprints on the transmitted light.
These absorption features trace structure in the universe at
epochs intermediate between the earliest density fluctuations seen in the
cosmic background radiation and the distribution of galaxies visible today.
The distribution of absorption features according to redshift ($z$) and the
column densities of gaseous material in different ions reveal the structure
of the IGM and its density and ionization state.  From the ionization state
of the gaseous species, we can also infer the processes responsible for
ionizing the gas, e.g., radiation from quasars in the early universe, or from
early bursts of star formation.

The lack of smooth \lya\ absorption by \hi\ in quasar spectra ({\it 1})
led to the conclusion that any diffusely distributed gas must be too highly
ionized to be visible.
Discrete absorption features were so plentiful, however, that they
were dubbed the ``Lyman $\alpha$ forest" ({\it 2}).
Because ionized helium (He$^+$) is more difficult to ionize than
hydrogen (ionization potential of 54.4 eV versus 13.6 eV) and recombines
faster, its abundance in the IGM is expected to be higher.  
Searches for a diffuse component of the IGM therefore concentrated
on absorption from \heii\ \lya\ $\lambda 304$ \AA\ ({\it 3}).
This absorption was first observed using the
{\it Hubble Space Telescope (HST)} ({\it 4})
along the line of sight to the quasar Q0302-003 ($z = 3.29$). The observation
showed the IGM to be essentially opaque at redshifts of $z \sim 3$, and
it seemed possible that the material in the \hi\ \lya\ forest
was sufficient to account for the \heii\ opacity ({\it 5}).
However, because no discrete \heii\ features could be observed, this required
several assumptions.
Measurement ({\it 6}) of the He$^+$ opacity at lower $z$,
using the {\it Hopkins Ultraviolet Telescope (HUT)}, showed a translucent
medium with \heii\ optical depth $\tau \sim 1$ at $z \sim 2.4$.
The \HUT\ observations were of insufficient spectral resolution to detect
discrete \heii\ absorption features.  However, models of the sightline towards
the quasar HS1700+64 ($z = 2.72$) required a substantial contribution to the
\heii\ opacity from He$^+$ that was more smoothly distributed than the
known \hi\ \lya\ absorbers ({\it 6, 7}) and improved
\HST\ observations of Q0302-003 also required more \heii\ opacity than
that predicted by the \hi\ absorption ({\it 8}).
Resolution of these inconsistencies requires observations at
higher spectral resolution.

Recent theoretical studies view the high-redshift IGM as a tracer of cosmic
structure formation by gravitational instability.  In such a scenario, diffusely
distributed baryonic material (the protons and neutrons that constitute ordinary
matter) responds to the gravitational influence of the underlying dark matter.
The most overdense regions collapse first to form the earliest galaxies. This
leaves the remaining gas as the IGM.
Rather than being a uniform medium filling the space between galaxies,
the IGM itself should show structur on scales larger than that of individual
galaxies.
Both numerical calculations ({\it 9--16}) and analytic theory ({\it 17--21})
link the evolving structure of the IGM to the \hi\
and \heii\ absorption visible in the spectra of high redshift quasars.

Given the unsettled observational issues and the need to test the
theoretical expectations, we planned {\it Far Ultraviolet Explorer (FUSE)}
observations of the \heii\ absorption toward the bright (visual magnitude
$\rm V = 16.1$) $z = 2.885$ quasar HE2347--4342 at high spectral resolution.
Previous observations of HE2347--4342 at longer ultraviolet (UV) wavelengths
showed it to be one of the brightest
candidates for such observations ({\it 22, 23}).
As with the other quasars observed with \HST, the \heii\ absorption in this
object is mostly opaque, but there are wavelength intervals of high
transmission that suggest we can see the beginnings of the He$^+$
re-ionization in the IGM in the redshift range accessible toward this
quasar ({\it 22}).
Because \heii\ \lya\ at $z = 2.885$ is redshifted to 1181 \AA\ and \HST\ is
sensitive only down to 1150 \AA, which corresponds to $z = 2.79$,
the \HST\ data do not cover a large range in redshift.
We therefore hoped to trace this re-ionization process over a greater
range with \FUSE\ because its short-wavelength sensitivity
would let us probe the IGM to much lower redshifts, down to $z \sim 2.0$
in principle ({\it 24}).

We observed HE2347--4342 with \FUSE\ in two separate campaigns.
The first observation, from
17 to 27 August 2000, comprised 351,672 s, of which
192,610 s was during orbital night ({\it 25}).
The second, running from 11 to 21 October 2000, accumulated 249,717 s,
with 183,630 s during orbital night.
For each observing campaign, HE2347--4342 was centered in the 30 arc sec
$\times$ 30 arc sec apertures.
To maintain the optical alignment of the four channels during these
campaigns, we offset \FUSE\ every other day to the nearby UV-bright star
WD2331--475 and performed routine adjustments to the mirror positions.

At the faint flux levels presented by HE2347--4342
($F_\lambda \sim 3 \times 10^{-15}~\rm erg~cm^{-2}~s^{-1}~\AA^{-1}$
at 1200 \AA), an accurate extraction
of a background-subtracted, calibrated source spectrum from the two-dimensional
(2D) recorded data required customized processing.
Because the LiF channels have a throughput that is $\sim$3 times that
of the SiC channels, we discuss only the LiF portions of the spectrum.
Because there are spatial variations in the background that affect the extracted
flux level in a source as faint as HE2347--4342, we turned off the standard
background subtraction in the \FUSE\ data-processing pipeline and extracted a
``source + background" spectrum ({\it 26}).
We then manually extracted background spectra from the geometrically rectified
detector images using regions lying adjacent to the source spectrum.
Dead spots and airglow lines from other \FUSE\ apertures were masked out
of these background spectra.
We computed a linear slope joining the two background regions, and scaled
the final result for each of the four detector segments to match the level at
the location of the extracted source+background spectrum.
We subtracted these background spectra from the pipeline-extracted
source+background spectra, and then used the rest of the normal pipeline
processes to perform the standard wavelength and flux calibrations.
At all steps we propagated a 1-$\sigma$ error array along with the data.

The multiplicity of detectors in \FUSE\ and our two separate observations
allowed several checks on our reduction process.
At each wavelength we obtained two separate extracted spectra from
each observation for a total of four independent spectra.
All spectral features noted in our combined spectrum were visible in
each independent spectrum.
As a check on our flux levels, we computed the scatter among the fluxes from
the 1000 to 1080 \AA\ spectra
($2.7 \times 10^{-16}~\rm erg~cm^{-2}~s^{-1}~\AA^{-1}$)
and the 1100 to 1187 \AA\ spectra
($2.1 \times 10^{-16}~\rm erg~cm^{-2}~s^{-1}~\AA^{-1}$).
For both, the scatter is higher than our 1-$\sigma$ errors,
indicating some residual systematic errors.
These errors are $\sim$5 to 10\% of the extrapolated continuum flux for
HE2347--4342, and they are indicative of our overall uncertainty.
As a check on our zero levels, we note that the interstellar absorption line
\cii\ $\lambda 1036$ is saturated in all \FUSE\ extragalactic spectra.
All four extracted spectra containing this line
have zero flux at line center to within the 1-$\sigma$ error bars.
At wavelengths of $>$1150 \AA,
previous \HST\ observations of HE2347--4342
show several regions where there is little or no flux.
Again, our extracted spectra also show no net flux in these regions
to within our 1-$\sigma$ errors.
Given the consistency among these separate observations, we combined the
separate spectra into a single spectrum with a uniform wavelength scale
and 0.05 \AA\ bins (Fig.~1).

Because the \FUSE\ bandpass stops short of wavelengths where the unobscured
continuum of HE2347--4342 is visible ($\lambda > 1190$ \AA), we obtained
low-resolution \HST\ spectra covering 1150--3200 \AA\ on 21 August
and 16 October 2000 to establish the continuum level.
These observations each consisted of a 1060 s exposure using
the Space Telescope Imaging Spectrograph (STIS) with grating G140L
and a 600 s exposure using grating G230L.
HE2347--4342 was slightly fainter (by 7\%) in the
October observation, but otherwise the spectra were identical.
We scaled the October observation up to the August flux levels, and
fitted a simple power law,
$F_\lambda = 3.31 \times 10^{-15} (\lambda/1000~\rm \AA)^{-2.40}~
\rm erg~cm^{-2}~s^{-1}~\AA^{-1}$,
with an extinction correction to spectral regions
free of galactic absorption lines.
We used a mean galactic extinction curve with ratio of selective to total
extinction $R_V = 3.1$ ({\it 27}) and a color excess $E(B - V) = 0.014$
({\it 28}), where $B$ is the blue-band magnitude.

The spectrum and extrapolated continuum (Fig.~1) shows that for $z > 2.72$
($\lambda \gtsim 1130$ \AA),
the IGM is relatively opaque, with occasional narrow windows that are nearly
transparent ({\it 22, 23}).
At $z=2.72$, the IGM becomes more transmissive, and the opacity systematically
drops at lower $z$ and shorter wavelengths.
In this translucent region, it is also apparent that discrete absorption
features account for most, if not all, of the opacity. We have resolved
the \heii\ absorption into a \heii\ \lya\ forest, analogous to the
\hi\ \lya\ forest.

The quantitative character of the evolution in opacity with redshift
is consistent with our expectations.
The mean opacity over the redshift interval 2.3 to 2.7 is
$\tau_{\rm HeII} = 0.91 \pm 0.01$ ({\it 29}), similar to the sightline
towards HS1700+64 ({\it 6}).
We compare the measured opacities to a model ({\it 21}) of discrete
clouds photoionized by the general quasar population, valid for epochs after
\heii\ re-ionization is complete.
In the model, ionization fluctuations on scales of
$\sim$4000 $\rm km~s^{-1}$ ($\sim$15 \AA, the size of our largest bins)
lead to the predicted range of opacities (Fig.~2).
For $z < 2.72$, where \heii\ reionization is
potentially complete, the absolute level of the observed opacity, its trend to
lower values at lower redshift, and the fluctuations about the mean are
all consistent with this model.

The extrapolated STIS continuum is consistent with the peak
fluxes in the \FUSE\ spectrum (Fig.~1).
Positive residuals from points above the extrapolated
continuum have a Gaussian distribution consistent with the 1$\sigma$
error bars on our data points, indicating that there are absorption-free
windows where the IGM is transparent in \heii\ \lya.
For models in which the physical density of smoothly distributed \heii\
evolves as $n_{\rm HeII} \propto (1 + z)^\alpha$,
and assuming a deceleration parameter $\Lambda = 0$,
and density parameter $\Omega = 1$,
the opacity in smoothly distributed \heii\ \lya\ varies with
$z$ as
$\tau_{\rm smooth}(z) = \tau_{\rm smooth}(z=2.885) [(1 + z) / (1 + 2.885)]^{\alpha-1.5}$
for $z < 2.885$ ({\it 1}).
By requiring that the distribution of the ratios of positive residuals to the
1-$\sigma$ errors be consistent with a Gaussian of mean 0 and dispersion 1,
we set a 1-$\sigma$ upper limit on $\tau_{\rm smooth}(z=2.885)$ by permitting
$\tau_{\rm smooth}(z=2.885)$, the power-law index,
and its normalization to vary freely until $\chi^2$ increases by 1.
This yields an upper limit of $\tau_{\rm smooth}(z=2.885) < 0.11$ to 0.12 for
$\alpha = 0$~to~6. 

Because the \FUSE\ data resolve the \heii\ \lya\ forest,
we can make a detailed comparison of the individual
absorption features to their counterparts in the \hi\ \lya\ forest.
To measure the individual \heii\ column densities, we fit the \FUSE\ spectrum
using the program SPECFIT ({\it 30}).
Our model incorporates the extrapolated continuum, individual
\heii\ absorption lines with Voigt profiles
(assuming Doppler widths of $b_{\rm HeII} = b_{\rm HI}$), and
foreground interstellar absorption in the Milky Way ({\it 31}).
For each \heii\ line, we permit the column density to vary, but we
fix $z$ and the line width ($b_{\rm HeII} = b_{\rm HI}$)
to have the same values as those in
the \hi\ line list from a Keck observation of HE2347--4342
with a resolution of $8~\rm km~s^{-1}$ ({\it 32}).
The \hi\ lines identified in the Keck spectrum
(184 in total) all have \heii\ counterparts (Fig.~3).
Although these lines account for most of the observed opacity in \heii\ \lya,
they account for only $\sim$50\% of the observed spectral features.
To model the remaining features,
we inserted additional lines whose redshifts were permitted
to vary freely and whose widths were fixed at $b = 27~\rm km~s^{-1}$,
the mean of the distribution from the \hi\ line list.
Over the 1000 to 1130 \AA\ wavelength range, 179 additional \heii\ \lya\ 
absorption features were added to our fit.
At the 3-$\sigma$ confidence limit, we are sensitive to \heii\ lines
with a limiting column density of $\rm log~N(HeII) > 10^{12.8}~cm^{-2}$.
For comparison, there are 72 \hi\ \lya\ features with
$\rm N_{HI} < 10^{13}~\rm cm^{-2}$ in the same corresponding range in $z$
in the Keck spectrum.

The added features are consistent with an extrapolation of the number of
\hi\ lines per unit column density of the form
$f(N_{HI}) \propto N_{HI}^{-1.5}$ ({\it 21, 33, 34})
down to column densities of $\sim 10^{11}~\rm cm^{-2}$.
This is potentially a substantial amount of material, but, given the
power-law form of the column density distribution, it does
not dominate the total mass in the Ly$\alpha$ forest ({\it 19}).
Because the physical sizes of the absorbers and the ionization
corrections are uncertain, we refrain from estimating
the amount of mass that this population of absorbers represents.
However, in the context of hydrodynamical models of the
IGM ({\it 14, 19}), we note that
the low-column-density extension of the \lya\ forest contains
$\sim$10\% of the baryons at $z=2$ to 3.
Our detection of a large number of \heii\ absorption features with
no \hi\ counterparts is consistent with the predictions of these models.

With measured ratios of \heii\ to \hi\ column densities,
$\eta = \rm N(He II)/N(H I)$, we can infer the shape of the ionizing
spectrum illuminating the absorbing gas.
Because our data include both measured values and lower limits (Fig.~4)
(for \heii\ \lya\ features with no \hi\ \lya\ counterparts), we
use the Kaplan-Meier estimator for censored data ({\it 35})
to derive a mean value $\langle \rm log~\eta \rangle = 1.89 \pm 0.04$
for the full data set.
Models of an IGM photoionized by
the integrated light from quasars propagated through
the IGM ({\it 21, 36}) predict values of $\eta = 30$ to 100
for quasar spectra with spectral indices $\alpha_q = 1.5$ to 2.1
($f_\nu \propto \nu^{-\alpha_q}$).
Intrinsic spectral indices for quasars (as measured down to a rest wavelength
of $\sim$350 \AA) lie in this range ({\it 37}).
Thus, most of the observed absorption features are consistent
with photoionization by quasar radiation.
However, about 40\% of the absorbers have
$\eta > 100$, indicating that localized regions of the
IGM are photoionized by softer spectra.  This might be
heavily filtered quasar radiation (in regions where the re-ionization
of \heii\ is not yet complete), or stellar radiation from
ongoing star formation~({\it 21})

There are fluctuations in $\eta$ on the scale of individual spectral features
(Fig.~4) which implies that the ionizing radiation field is not uniform.
Local radiation sources
(which may be either quasars or star-forming galaxies)
probably affect $\eta$ and the local opacity of the IGM.
This characteristic is most evident
at high $z$ in the low-opacity ``windows" first seen in \HST\ spectra
of HE2347--4342 ({\it 22, 23}), but it persists to lower
redshift where the IGM is less opaque.

By resolving the \heii\ \lya\ absorption in the IGM using \FUSE, we are able
to see structure in the universe that extends into the lowest density
regions. The presence of \heii\ absorption with no \hi\ counterparts is
consistent with hydrodynamical models that predict density fluctuations
due to gravitational instabilities on all scales, from the high density peaks
that form galaxies to the distribution of gas in low-density voids
({\it 9--16, 19, 20}).
Observations of \heii\ \lya\ absorption is the best method to study these
regions, and future observations of additional bright quasars with \FUSE\ should
provide essential information on the cosmic variance in the structures we see
along different lines of sight.

\newpage
\onecolumn

\begin{figure}
\plotfiddle{"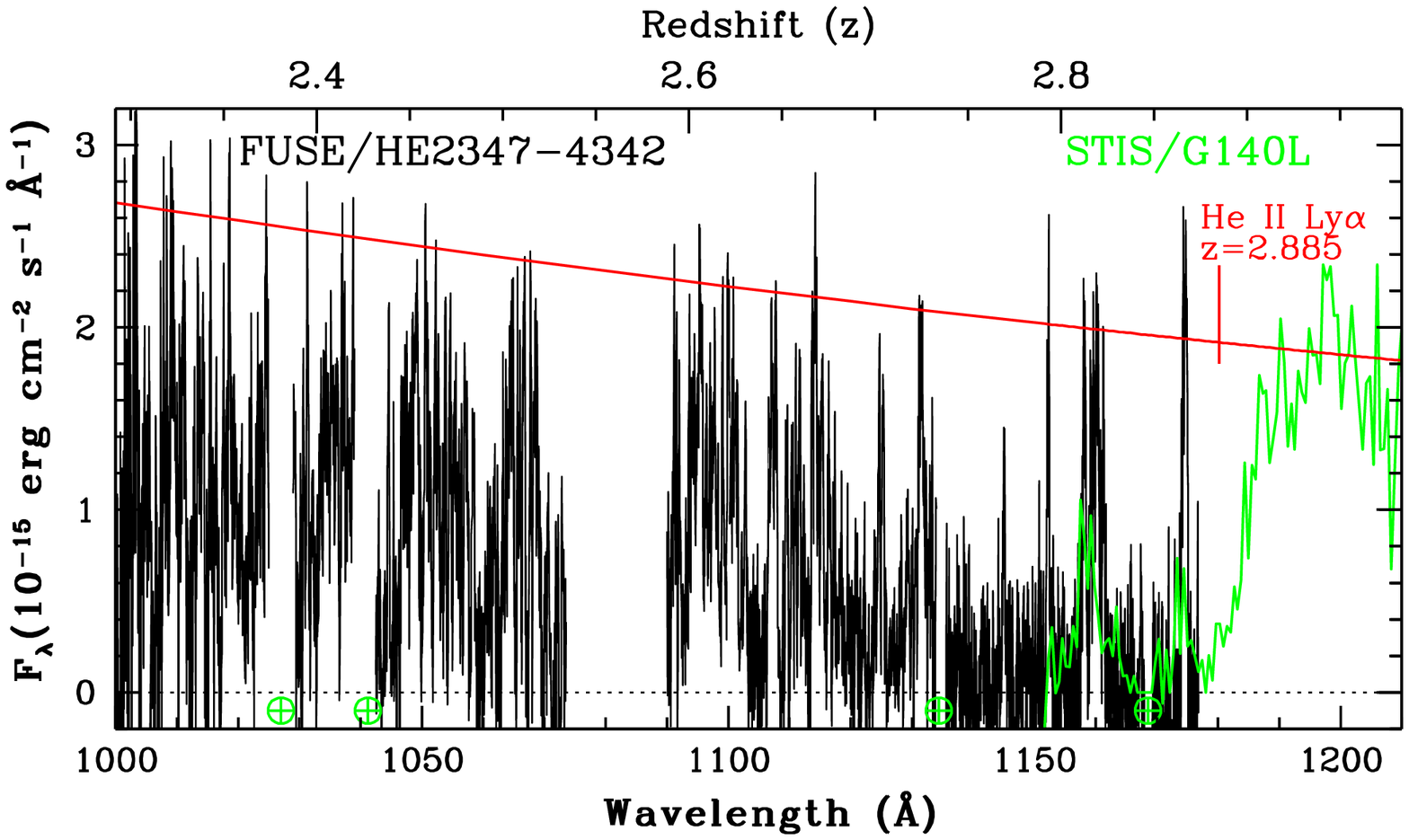"}{4.55 in} {0}{90}{90}{-275}{-100}
\caption{
\FUSE\ spectrum of HE2347--4342 (binned to 0.05 \AA\ pixels) is shown in
black.  Bins at the peak fluxes have a signal-to-noise ratio of $\sim$7.
A portion of the contemporaneous STIS spectrum is shown in green.
The red line is the extrapolation of the power law plus extinction
of E(B-V)=0.014 fitted to the STIS spectrum.
The position of He~II Ly$\alpha$ at the redshift of HE2347--4342 is marked.
Gaps in the \FUSE\ spectrum due to excised terrestrial airglow lines are
marked with green $\oplus$.
The broad gap from 1072 to 1089 \AA\ is due to gaps between the \FUSE\ detector
segments.
\label{he2347f1.ps}}
\end{figure}

\begin{figure}
\plotfiddle{"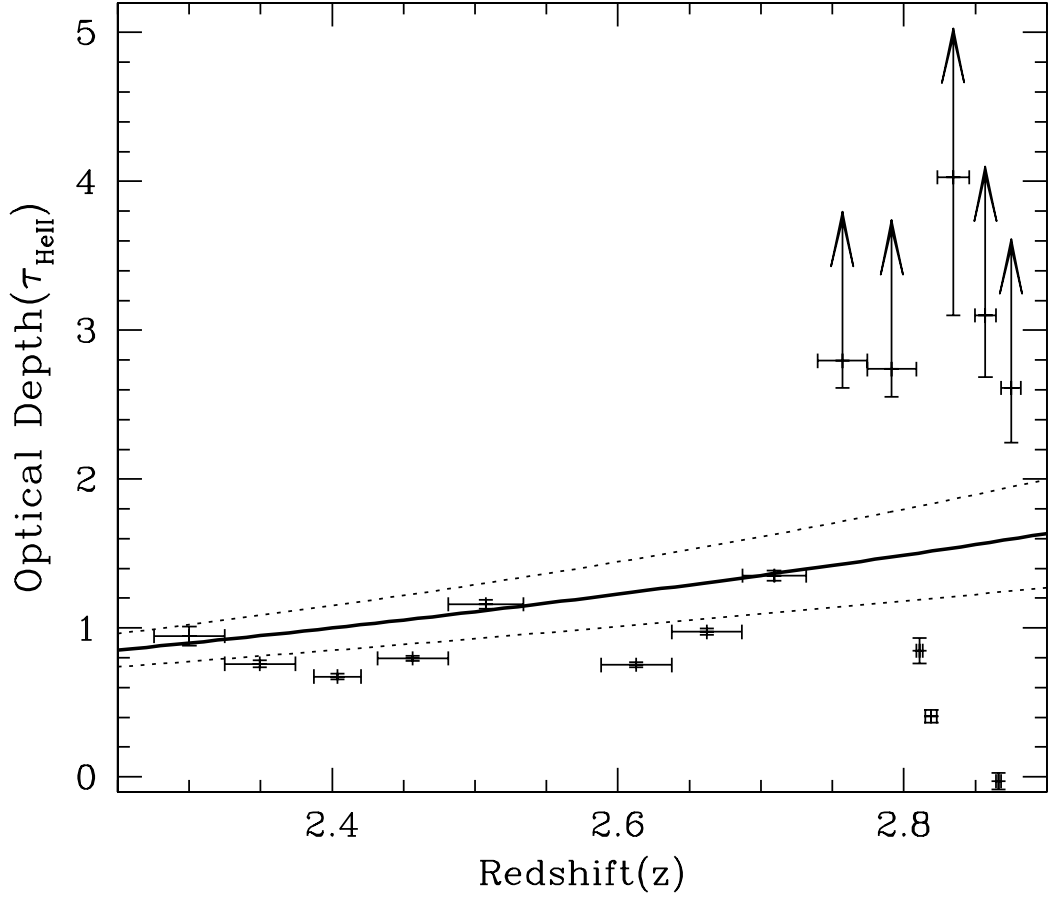"}{4.55 in} {0}{100}{100}{-305}{-250}
\caption{
The He~II opacity $\tau_{HeII}$ in coarse 5 to 15 \AA\ bins is shown as a
function of redshift.
The solid curve is the theoretical prediction
for the opacity assuming line blanketing only ({\it 21}).
The dashed curves indicate the 1 $\sigma$ extent
of variations in the opacity anticipated due to ionization fluctuations.
At redshifts below $z =2.72$, where the reionization of He II in the IGM
appears to be complete, the model matches well the observed opacity, its
trend to lower values at lower redshift, and the fluctuations about the mean.
\label{he2347f2.ps}}
\end{figure}

\begin{figure}
\plotfiddle{"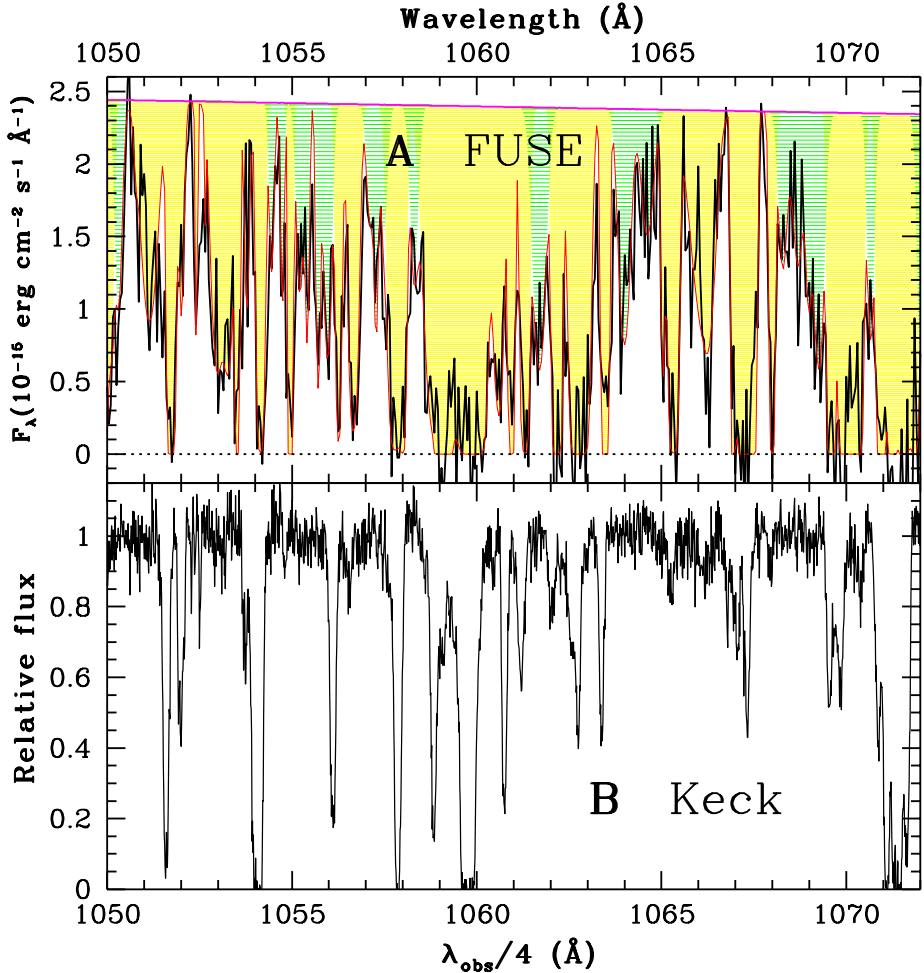"}{4.55 in} {0}{70}{70}{-212}{-100}
\caption{
({\bf A})
The black line is a portion of the \FUSE\ spectrum of HE2347--4342.
The smooth red curve across the top is the extrapolated continuum.
The magenta line overlaying the spectral data in black is the model
described in the text.  The area shaded in yellow shows the fraction of
the He~II opacity due to absorption features that correspond to H~I absorption
lines identified in the Keck spectrum ({\it 32}).
The area shaded in green shows the
fraction of the opacity due to additional He~II absorption features that have
no H~I counterparts in the Keck spectrum.
({\bf B})
The normalized Keck spectrum ({\it 32}).
Note the direct correspondence between the wavelengths of H~I lines in the Keck
spectrum and He~II absorption features in the \FUSE\ spectrum.
\label{he2347f3.ps}}
\end{figure}

\begin{figure}
\plotfiddle{"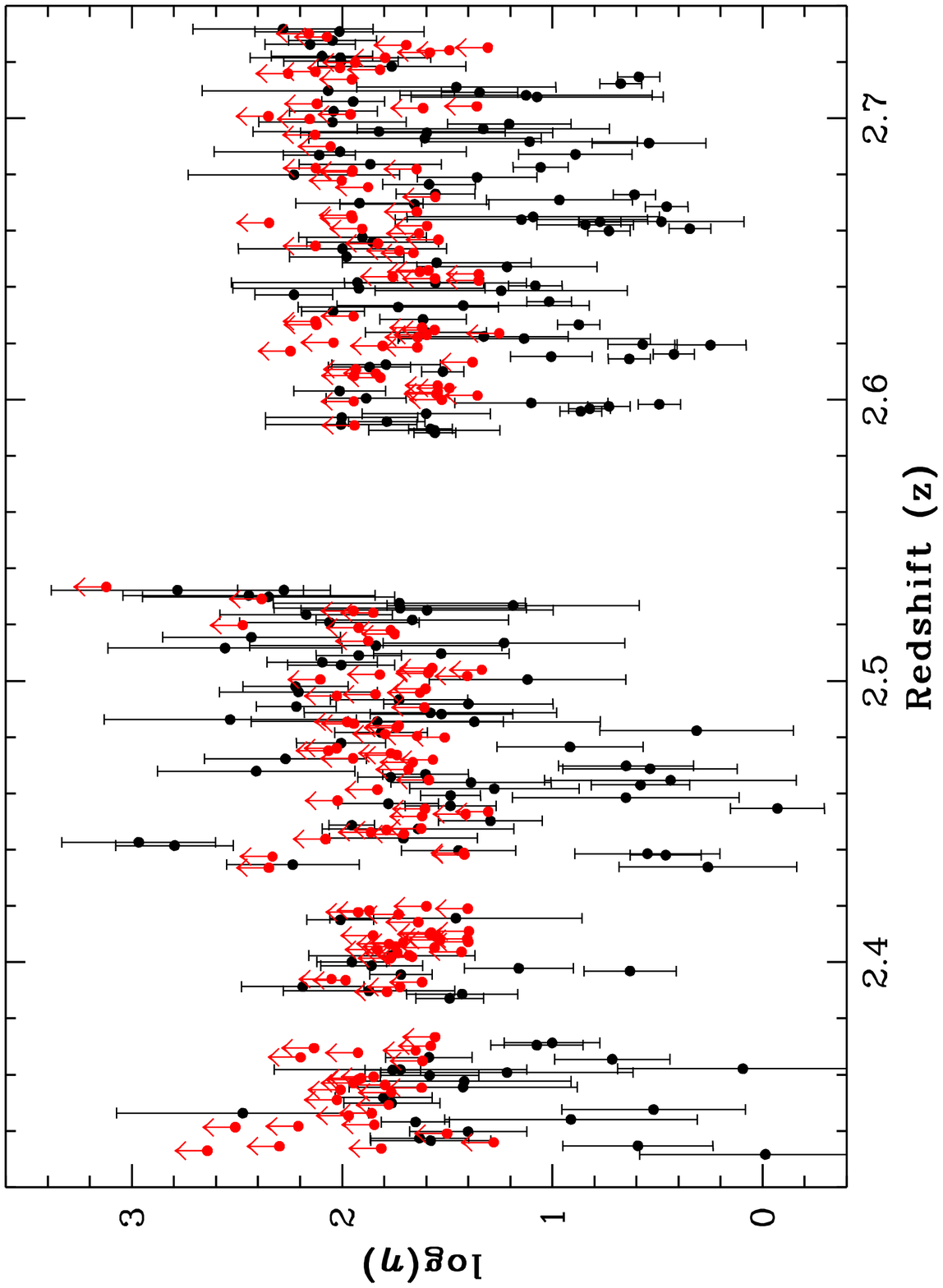"}{4.55 in} {-90}{70}{70}{-275}{420}
\caption{
The logarithm of the He~II to H~I column density ratio $\eta$=N(HeII)/N(HI)
versus redshift for measured absorption features in the \FUSE\ spectrum.
The black points with error bars have measured column densities for both
HeII (from \FUSE) and H~I (from Keck).
The red points are lower limits computed using the He~II column density
measured from the \FUSE\ spectrum and an upper limit
of $10^{12.3}~\rm cm^{-2}$ for H~I absorption lines in the Keck spectrum.
A Spearman rank correlation coefficient of 0.05 indicates that there is
no significant trend with redshift.
The mean value, $\langle \log \eta \rangle = 1.89$, is
consistent with photoionization of the IGM by integrated quasar light
({\it 21, 36}).
However, one can see significant fluctuations in $\eta$ that indicate
that the ionizing radiation field is not uniform.
Values of $\eta > 100$ indicate regions photoionized by either
heavily filtered quasar radiation or bursts of star formation.
\label{he2347f4.ps}}
\end{figure}


\begin{thebibliography}{}

\bibitem[{\it 1}]{GP65}
	1. J. Gunn and B. Peterson, {\it Astrophys.J. \bf 142}, 1633 (1965).
\bibitem[{\it 2}]{Lynds71}
	2. R. Lynds, {\it Astrophys.J.} {\bf 164}, L73 (1971).
\bibitem[{\it 3}]{Jakobsen93}
	3. P. Jakobsen, et al., {\it Astrophys.J. \bf 417}, 528 (1993).
\bibitem[{\it 4}]{Jakobsen94}
	4. P. Jakobsen, et al., {\it Nature \bf 370}, 35 (1994).
\bibitem[{\it 5}]{Songaila95}
	5. A. Songaila, E. M. Hu, L. L. Cowie, {\it Nature \bf 375}, 124
	(1995).
\bibitem[{\it 6}]{Davidsen96}
	6. A. F. Davidsen, G. A. Kriss, W. Zheng, {\it Nature \bf 380}, 47
	(1996).
\bibitem[{\it 7}]{Zheng98}
	7. W. Zheng, A. F. Davidsen, G. A. Kriss,
	{\it Astron.J. \bf 115}, 391 (1998).
\bibitem[{\it 8}]{Heap2000}
	8. S. R. Heap, et al., {\it Astrophys.J. \bf 534}, 69 (2000).
\bibitem[{\it 9}]{Cen94}
	9. R. Cen, J. Miralda-Escud\'e, J. P. Ostriker, M. Rauch,
	{\it Astrophys.J. \bf 437}, L9 (1994).
\bibitem[{\it 10}]{Zhang95}
	10. Y. Zhang, P. Anninos, M. L. Norman,
	{\it Astrophys.J. \bf 453}, L57 (1995).
\bibitem[{\it 11}]{Zhang97}
	11. Y. Zhang, P. Anninos, M. L. Norman, A. Meiksin,
	{\it Astrophys.J. \bf 485}, 496 (1997).
\bibitem[{\it 12}]{Hernquist96}
	12. L. Hernquist, N. Katz, D. H. Weinberg, J. Miralda-Escud\'e,
	{\it Astrophys.J. \bf 457}, L51 (1996).
\bibitem[{\it 13}]{ME96}
	13. J. Miralda-Escud\'e, R. Cen, J. P. Ostriker M. Rauch,
	{\it Astrophys.J \bf 471}, 582 (1996).
\bibitem[{\it 14}]{Croft97}
	14. R. A. C. Croft, D. H. Weinberg, N. Katz, L. Hernquist,
	{\it Astrophys.J. \bf 488}, 532 (1997).
\bibitem[{\it 15}]{Cen99}
	15. R. Cen, and J. P. Ostriker, {\it Astrophys.J. \bf 514}, 1 (1999).
\bibitem[{\it 16}]{Dave99}
	16. R. Dav\'e, L. Hernquist, N. Katz, D. H. Weinberg,
	{\it Astrophys.J. \bf 511}, 521 (1999).
\bibitem[{\it 17}]{Bi93}
	17. H. Bi, {\it Astrophys.J. \bf 405}, 479 (1993).
\bibitem[{\it 18}]{Bi95}
	18. H. Bi, J. Ge, L.-Z. Fang, {\it Astrophys.J. \bf 452}, 90 (1995).
\bibitem[{\it 19}]{Bi97}
	19. H. Bi and A. F. Davidsen, {\it Astrophys.J. \bf 479}, 523 (1997).
\bibitem[{\it 20}]{Hui97}
	20. L. Hui, N. Y. Gnedin, Y. Zhang, {\it Astrophys.J. \bf 486},
	599 (1997).
\bibitem[{\it 21}]{Fardal98}
	21. M. A. Fardal, M. L. Giroux, J. M. Shull,
	{\it Astron.J. \bf 115}, 2206 (1998).
\bibitem[{\it 22}]{Reimers97}
	22. D. Reimers, et al., {\it Astron.Astrophys. \bf 327}, 890 (1997).
\bibitem[{\it 23}]{Smette01}
	23. A. Smette et al. {\it Astrophys.J.}, in press
	(available at http://xxx.lanl.gov/abs/astro-ph/0012193).
\bibitem[{\it 24}]{Footnote1}
24. \FUSE\ covers the wavelength range 912 to 1187 \AA\
with four separate optical channels ({\it 38, 39}).
Each channel consists of an off-axis paraboloidal mirror feeding a
prime-focus Rowland circle spectrograph.
The dispersed light is focused on two, 2D photon-counting
microchannel-plate detectors with KBr photocathodes that record the
time, position, and pulse height of each photon event.
Each detector is split into two segments with a small gap between them.
Two channels use LiF-coated optics to cover the band
1000 to 1187 \AA;
the other two channels use SiC coatings to provide
wavelength coverage down to 912 \AA. 

\bibitem[{\it 25}]{Footnote2}
25. Because the count rate in a 50,000 s observation in June 2000 was
lower than the typical background rate,
we chose the dates of these observations to maximize
the exposure time during orbital night, the part of the orbit when \FUSE\ is
in Earth's shadow.  This minimizes the background contribution
due to scattered geocoronal \lya.

\bibitem[{\it 26}]{Footnote3}
	26. To obtain the lowest background rate possible,
	we used only orbital night observations, filtered the high-background
	``bursts" ({\it 39}) from the data, and
	used only events with pulse heights in the range 4 to 16.
	This reduces noise due to the low pulse-height events that arise from
	internal detector background rather than from real photons.

\bibitem[{\it 27}]{Cardelli89}
	27. J. Cardelli, G. Clayton, J. Mathis,
	{\it Astrophys.J. \bf 345}, 245 (1989).
\bibitem[{\it 28}]{Schlegel98}
	28. D. J. Schlegel, D. P. Finkbeiner, M. Davis,
	{\it Astrophys.J. \bf 500}, 525 (1998).
\bibitem[{\it 29}]{Footnote4}
	29. Although the mean opacity is tightly constrained, we note that
	there is a large dispersion of 0.9 about this value.
\bibitem[{\it 30}]{Kriss94}
        30. G. A. Kriss, in {\it Astronomical Data Analysis Software and
	Systems III}, D. R. Crabtree, R. J.  Hanisch, J. Barnes, Eds.
	({\it A. S. P. Conf. Series 61}, ASP, San Francisco, 1994), p. 437.
\bibitem[{\it 31}]{Footnote5}
	31. For the low extinction along this sightline, the interstellar
	absorption model contributes strong features only for
	Si~{\sc ii} $\lambda 1020$, Ly$\beta$ $\lambda 1025$,
	\cii\ $\lambda 1036$, {\sc O~i} $\lambda 1039$,
	Ar~{\sc i} $\lambda\lambda 1048,1066$,
	and no $\rm H_2$ absorption.
	Along similar high latitude sight lines observed with
	\FUSE, the column density in $\rm H_2$ is on the order of a few
	$\times 10^{15}~\rm cm^{-2}$
	[J. M. Shull et al., {\it Astrophys.J. \bf 538}, L73 (2000)].
	The average opacity produced by such a column density amounts to
	$< 4$\% from 1000--1110~\AA.
\bibitem[{\it 32}]{Songaila98}
	32. A. Songaila, {\it Astron.J. \bf 115}, 2184 (1998).
\bibitem[{\it 33}]{PR93}
	33. W. H. Press and G. B. Rybicki,{\it Astrophys.J. \bf 418}, 585 (1993).
\bibitem[{\it 34}]{Hu95}
	34. E. M. Hu, T.-S. Kim, L. L. Cowie, A. Songaila, M. Rauch,
	{\it Astron.J. \bf 110}, 1526 (1995).
\bibitem[{\it 35}]{Feigelson85}
	35. E. D. Feigelson, and P. I. Nelson, {\it Astrophys.J. \bf 293},
	192 (1985).
\bibitem[{\it 36}]{Madau99}
	36. P. Madau, F. Haardt, M. J. Rees,
	{\it Astrophys.J. \bf 514}, 648 (1999).
\bibitem[{\it 37}]{Zheng97}
	37. W. Zheng, G. A. Kriss, R. C. Telfer, J. P. Grimes, A. F. Davidsen,
	{\it Astrophys.J. \bf 475}, 469 (1997).
\bibitem[{\it 38}]{Moos2000}
	38. H. W. Moos, et al., {\it Astrophys.J. \bf 538}, L1 (2000).
\bibitem[{\it 39}]{Sahnow2000}
	39. D. Sahnow, et al., {\it Astrophys.J. \bf 538}, L7 (2000).

\bibitem[{\it 40}]{Ack}
	40.  We sadly report that Arthur F. Davidsen, one of our key
	team members and a pioneer in ultraviolet observations of the
	intergalactic medium, passed away on July 19, 2001.
	This work is based on data obtained for the Guaranteed Time Team by
	the NASA-CNES-CSA \FUSE\ mission operated by the Johns Hopkins 
	University. Financial support to U. S. participants has been provided by
	NASA contract NAS5-32985, and by the NASA LTSA grant NAG5-7262
	to the University of Colorado.
	A portion of this work is based on observations with the NASA/ESA
	{\it Hubble Space Telescope}, obtained at the Space Telescope Science
	Institute, which is operated by the Association of Universities for
	Research in Astronomy, Inc., under NASA contract NAS5-26555.
	These observations are associated with proposal ID 8875.
	We thank B. Roberts for his efforts in planning and scheduling
	the successful \FUSE\ observations, and J. Kim for her help in
	analyzing the STIS data.

\end{thebibliography}
\end{document}